# Interpretation of measured 3x3 partial depolarizing Mueller matrices

José J. Gil[1,*], Ignacio San José[2], Mónica Canabal-Carbia[3], Juan Campos[3], Angel Lizana[3], Irene Estévez[3]

[1] *Group of Photonic Technologies, Universidad de Zaragoza, Pedro Cerbuna 12, 50009 Zaragoza, Spain*

[2] *Instituto Aragonés de Estadística, Gobierno de Aragón, Bernardino Ramazzini 5, 50015 Zaragoza, Spain*

[3] *Grup d'Òptica, Departament de Física, Universitat Autònoma de Barcelona, 08193 Bellaterra, Spain*

[*] ppgil@unizar.es

Mueller polarimetry is a powerful technique with broad applications in astronomy, remote sensing, advanced material analysis, and biomedical imaging. However, instrumental constraints frequently restrict the measurement to an incomplete Mueller matrix limited to its upper-left 3×3 submatrix. Simply padding the missing entries with zeros to form a 4×4 matrix can produce physically inconsistent results, even for nondepolarizing systems. To address this issue, we present a systematic procedure to complete 3×3 measured Mueller matrices into physically consistent 4×4 matrices. The method relies on the covariance matrix formalism and selects, among the infinitely many admissible completions, the one with maximal polarimetric purity. This criterion ensures that the synthesized matrix corresponds to the least random (most deterministic) model compatible with the measurement. The procedure is fully general and can be applied to any 3×3 partial Mueller polarimetric data, providing a reliable and physically grounded reconstruction tool for polarimetric imaging and materials characterization.

## 1. Introduction

Although the complete Mueller matrix provides full information on the polarimetric behavior of a sample for a given measurement configuration (defined by factors such as the spectral characteristics of the probing light, the incidence or scattering geometry, the illuminated area, and the sample's physical conditions), practical instruments often yield only partial measurements, typically restricted to linear polarization states [1].

For instance, division-of-focal-plane or single-shot polarimetric cameras, now widely used in various applications [2-5], rely on microgrid polarizer arrays that enable simultaneous analysis of the linear polarization content of the incident light. Similarly, polarimeters based on conical refraction [6] often provide only partial Mueller matrix information, typically limited to the upper-left 3×3 submatrix. Other types of incomplete polarimetric measurements may instead lack only one row or one column, depending on the instrument architecture and measurement constraints. Such limitations stem from practical constraints including compactness, acquisition speed, and restricted illumination power, which impose a trade-off between measurement completeness and experimental feasibility.

A particularly relevant case is found in the development of Mueller polarimeters designed for medical imaging, where such instruments have become powerful tools for tissue inspection, analysis, and diagnosis [7-15]. For endoscopic systems, the design often requires restricting the probing states of polarization to linear ones, as a compromise between acquisition speed, signal-to-noise ratio, and illumination safety.

These examples highlight a common trade-off in polarimetric system design, where the completeness of the Mueller matrix analysis is often constrained by the practical limitations of the instrumentation and restricted to incomplete polarimetry.

While recent approaches have attempted to address this problem using machine-learning algorithms trained on fully characterized polarimetric data [16], a physically grounded method capable of reconstructing complete Mueller matrices from partial measurements without relying on a priori sample information remains to be developed. Therefore, a procedure to implement a compatible and appropriate full 4×4 Mueller matrix from partial data would be highly valuable in this context.

In this work, we address the specific case of 3×3 measured matrices and propose a procedure for synthesizing a physically consistent 4×4 Mueller matrix that incorporates the available information while fulfilling natural constraints. These include compatibility with the measured data, consistency with physical requirements, and maximal polarimetric purity.

Filling unmeasured elements with zeros is generally not a suitable option. A good example of this is the case of a material sample with deterministic polarimetric behavior or exhibiting certain symmetries, whose filling procedure is based on very specific strategies [17,18].

The general approach presented is grounded on the properties of the covariance matrix associated with a Mueller matrix. In particular, it takes advantage of the fact that the real part of this covariance matrix is determined (up to a single unknown parameter) by the upper-left 3×3 submatrix provided by the incomplete polarimetric measurement.

To develop this approach, the paper is organized as follows: Section 2 introduces the necessary concepts and notations. Section 3 establishes the criteria that govern the completion procedure, which is fully described in Section 4. Section 5 is devoted to the application of the procedure to some illustrative examples.

## 2. Theoretical background

The transformation of the Stokes vectors upon linear interaction with a medium can be represented by a Mueller matrix $\mathbf{M}$ that, applied to the Stokes vector $\mathbf{s}$ of the incident light beam, leads to the Stokes vector of the emerging light $\mathbf{s}' = \mathbf{M}\mathbf{s}$.

The elements of $\mathbf{M}$ are denoted as $m_{ij}$ $(i,j = 0,1,2,3)$. For convenience, $\mathbf{M}$ can be written in block form as [19,20]

$$\mathbf{M} = m_{00} \begin{pmatrix} 1 & \mathbf{D}^{\mathrm{T}} \\ \mathbf{P} & \mathbf{m} \end{pmatrix}, \tag{1}$$

where superscript T denotes transpose; $m_{00}$ is the mean intensity attenuation coefficient; $\mathbf{D}$ and $\mathbf{P}$ are the diattenuation and polarizance vectors, whose respective absolute values are the diattenuation $D$ and polarizance $P$, and the submatrix $\mathbf{m}$, with elements $m_{kl}/m_{00}$ $(k,l = 1,2,3)$, whose Frobenius norm is

$$\|\mathbf{m}\| = \frac{1}{m_{00}}\sqrt{\sum_{k,l=1}^{3} m_{kl}^2} = \sqrt{\mathrm{tr}(\mathbf{m}^{\mathrm{T}}\mathbf{m})} = \sqrt{3}P_S, \tag{2}$$

$P_S$ being the polarimetric dimension index (also called degree of spherical purity), which is bounded by $0 \le P_S \le 1$ [21,22].

The degree of polarimetric purity, or depolarization index, is defined as [23]

$$P_\Delta = \sqrt{\frac{\mathrm{tr}\,\mathbf{M}^{\mathrm{T}}\mathbf{M} - m_{00}^2}{3 m_{00}^2}} = \sqrt{\frac{D^2 + P^2 + P_S^2}{3}}, \tag{3}$$

with $0 \le P_\Delta \le 1$. Mueller matrices for which $P_\Delta = 1$ (i.e., preserving the degree of polarization of incident totally polarized light) are called pure, also nondepolarizing or Mueller-Jones matrices. When $P_\Delta < 1$, the Mueller matrix is called nonpure, or depolarizing. Pure Mueller matrices can be derived from Jones matrices and thus inherits a peculiar mathematical structure that depends on up to seven independent parameters [24]. In addition, natural deterministic (nondepolarizing) interactions do not amplify the intensity of light and consequently $m_{00}$ satisfy the passivity condition $m_{00} \le 1/(1+D)$ [26,27] (recall that diattenuation and polarizance are equal for pure Mueller matrices [25]).

Since a general Mueller matrix summarizes an integral polarimetric behavior derived from a temporal, spatial and spectral average of a number of elementary interactions (represented by respective pure Mueller matrices) [24-35] the structure and properties of depolarizing Mueller matrices relies the fact that they can always be expressed as an average (convex sum) of pure (and passive) Mueller matrices. As a consequence, any physical Mueller matrix depends on up to sixteen independent parameters and must satisfy two types of inequalities, namely the passivity condition $m_{00} \le 1/(1+Q)$, with $Q = \max(D,P)$ [27,36], and the four covariance conditions consisting of the nonnegativity of the four eigenvalues of the (Hermitian) covariance matrix $\mathbf{H}$ associated with $\mathbf{M}$ and defined below [28,37]. For the purposes of the present work, only the covariance conditions will be considered, since they are sufficient to ensure the physical consistency required for the proposed reconstruction procedure. The passivity condition, although relevant for general analyses of optical systems, can be omitted here without affecting the validity or applicability of the method.

The explicit expression of $\mathbf{H}$ in terms of the elements of $\mathbf{M}$ is [28,37]

$$\mathbf{H}(\mathbf{M}) =$$

$$\frac{1}{4}\begin{pmatrix} m_{00}+m_{01} & m_{02}+m_{12} & m_{20}+m_{21} & m_{22}+m_{33} \\ +m_{10}+m_{11} & +i(m_{03}+m_{13}) & -i(m_{30}+m_{31}) & +i(m_{23}-m_{32}) \\ m_{02}+m_{12} & m_{00}-m_{01} & m_{22}-m_{33} & m_{20}-m_{21} \\ -i(m_{03}+m_{13}) & +m_{10}-m_{11} & -i(m_{23}+m_{32}) & -i(m_{30}-m_{31}) \\ m_{20}+m_{21} & m_{22}-m_{33} & m_{00}+m_{01} & m_{02}-m_{12} \\ +i(m_{30}+m_{31}) & +i(m_{23}+m_{32}) & -m_{10}-m_{11} & +i(m_{03}-m_{13}) \\ m_{22}+m_{33} & m_{20}-m_{21} & m_{02}-m_{12} & m_{00}-m_{01} \\ -i(m_{23}-m_{32}) & +i(m_{30}-m_{31}) & -i(m_{03}-m_{13}) & -m_{10}+m_{11} \end{pmatrix}. \tag{4}$$

Conversely,

$$\mathbf{M}(\mathbf{H}) =$$

$$\begin{pmatrix} h_{00}+h_{11} & h_{00}-h_{11} & h_{01}+h_{10} & -i(h_{01}-h_{10}) \\ +h_{22}+h_{33} & +h_{22}-h_{33} & +h_{23}+h_{32} & -i(h_{23}-h_{32}) \\ h_{00}+h_{11} & h_{00}-h_{11} & h_{01}+h_{10} & -i(h_{01}-h_{10}) \\ -h_{22}-h_{33} & -h_{22}+h_{33} & -h_{23}-h_{32} & +i(h_{23}-h_{32}) \\ h_{02}+h_{20} & h_{02}+h_{20} & h_{03}+h_{30} & -i(h_{03}-h_{30}) \\ +h_{13}+h_{31} & -h_{13}-h_{31} & +h_{12}+h_{21} & +i(h_{12}-h_{21}) \\ i(h_{02}-h_{20}) & i(h_{02}-h_{20}) & i(h_{03}-h_{30}) & h_{03}+h_{30} \\ +i(h_{13}-h_{31}) & -i(h_{13}-h_{31}) & +i(h_{12}-h_{21}) & -h_{12}-h_{21} \end{pmatrix}. \tag{5}$$

The real part of $\mathbf{H}$, depends on the nine elements $m_{ij}$ $(i,j = 0,1,2)$ from the upper left 3×3 submatrix of $\mathbf{M}$, plus $m_{33}$, while the imaginary part depends on the elements of the last file and column of $\mathbf{M}$, excluding $m_{33}$.

$\mathbf{H}$ is fully characterized by its eigenvalue-eigenvector structure, $\mathbf{H} = \mathbf{U}\boldsymbol{\Lambda}\mathbf{U}^\dagger$, where the dagger stands for conjugate transpose, $\mathbf{U}$ is the unitary matrix whose columns $(\mathbf{u}_0, \mathbf{u}_1, \mathbf{u}_2, \mathbf{u}_3)$ are the eigenvectors of $\mathbf{H}$ (which correspond to respective pure Mueller matrices), and $\boldsymbol{\Lambda} = \mathrm{diag}(\lambda_0, \lambda_1, \lambda_2, \lambda_3)$, where, as in other related papers, the eigenvalues are taken so as $\lambda_0 \ge \lambda_1 \ge \lambda_2 \ge \lambda_3 \ge 0$. Consequently, $\mathbf{H}$ can be expressed as a convex sum of up to four statistically pure covariance matrices [28,37]:

$$\mathbf{H} = \lambda_0 \hat{\mathbf{H}}_{J0} + \lambda_1 \hat{\mathbf{H}}_{J1} + \lambda_2 \hat{\mathbf{H}}_{J2} + \lambda_3 \hat{\mathbf{H}}_{J3}, \tag{6}$$

$$\left[\lambda_0 \ge \lambda_1 \ge \lambda_2 \ge \lambda_3 \ge 0, \ \hat{\mathbf{H}}_{Ji} = \mathbf{u}_i \otimes \mathbf{u}_i^\dagger \ (i=0,1,2,3)\right],$$

where $\otimes$ represents the Kronecker product and the subscript $J$ indicates that the matrix is statistically pure.

The above spectral decomposition can be expressed in terms of up to four pure Mueller as follows

$$\mathbf{M} = \lambda_0 \hat{\mathbf{M}}_{J0} + \lambda_1 \hat{\mathbf{M}}_{J1} + \lambda_2 \hat{\mathbf{M}}_{J2} + \lambda_3 \hat{\mathbf{M}}_{J3}, \tag{7}$$

where the particular form of each normalized pure Mueller matrix $\hat{\mathbf{M}}_{Ji}$ derived from the corresponding unit eigenvector $\mathbf{u}_i$, which in turn determines the associated 2×2 normalized complex Jones matrix $\mathbf{T}_i$.

Since an unphysical global phase factor of $\mathbf{T}_i$ is lost when it is transformed to $\hat{\mathbf{M}}_{Ji}$, it depends on up to six independent parameters (the seventh free parameter being $\lambda_i$). Thus, the orthonormality among the eigenvectors of $\mathbf{H}$ implies that, once $\mathbf{u}_0$ is determined from its six parameters, then $\mathbf{u}_1$ (orthonormal to $\mathbf{u}_0$) depends on up to four extra independent parameters, then $\mathbf{u}_2$ depends on two extra parameters, while $\mathbf{u}_3$ is fixed from the other three eigenvectors.

Consequently, a single-component system, $\mathbf{M} = \lambda_0 \hat{\mathbf{M}}_{J0}$ (rank $\mathbf{H} = 1$), depends on up to seven free parameters (one from $\lambda_0$ and six from $\mathbf{u}_0$); a two-component system (rank $\mathbf{H} = 2$) depends on up to twelve free parameters (seven from the first component, one from $\lambda_1$, and four from $\mathbf{u}_1$); a three-component system (rank $\mathbf{H} = 3$) depends on up to fifteen free parameters (twelve from the two first components, one from $\lambda_2$ and two from $\mathbf{u}_2$); and, as expected, a four-component system depends on up to sixteen free parameters.

The spectral decomposition of $\mathbf{H}$ can be transformed to the following characteristic decomposition [38]

$$\mathbf{H} = m_{00}\left[ P_1 \hat{\mathbf{H}}_{J0} + (P_2 - P_1)\hat{\mathbf{H}}_1 + (P_3 - P_2)\hat{\mathbf{H}}_2 + (1 - P_3)\hat{\mathbf{H}}_3 \right]$$

$$\begin{bmatrix} \hat{\mathbf{H}}_{J0} = \mathbf{u}_0 \otimes \mathbf{u}_0^\dagger, \quad \hat{\mathbf{H}}_1 = \frac{1}{2}\sum_{i=0}^{1} \mathbf{u}_i \otimes \mathbf{u}_i^\dagger, \\ \hat{\mathbf{H}}_2 = \frac{1}{3}\sum_{i=0}^{2} \mathbf{u}_i \otimes \mathbf{u}_i^\dagger, \quad \hat{\mathbf{H}}_3 = \frac{1}{4}\sum_{i=0}^{3} \mathbf{u}_i \otimes \mathbf{u}_i^\dagger = \frac{1}{4}\mathbf{I}, \end{bmatrix} \quad (8)$$

where $\mathbf{I}$ is the identity matrix, $\operatorname{tr}\mathbf{H} = m_{00}$, and the coefficients affecting the components are governed by the three indices of polarimetric purity (IPP) [39],

$$P_1 = \frac{\lambda_0 - \lambda_1}{m_{00}}, \quad P_2 = \frac{\lambda_0 + \lambda_1 - 2\lambda_2}{m_{00}},$$
$$P_3 = \frac{\lambda_0 + \lambda_1 + \lambda_2 - 3\lambda_3}{m_{00}}, \quad (9)$$

which provide, in a scaled manner ($0 \leq P_1 \leq P_2 \leq P_3 \leq 1$) [39], complete information on the structure of polarimetric purity-randomness of $\mathbf{M}$.

This decomposition can be expressed in terms of Mueller matrices by replacing $\mathbf{H}$ by $\mathbf{M}$.

### 3. Criteria for completing the Mueller matrix

When partial Mueller polarimetry yields a measured 3×3 matrix, it provides nine known parameters out of the full set of sixteen that define a complete of $\mathbf{M}$. The remaining seven parameters (corresponding to the last row and column), are unknown and, except for pure Mueller matrices and certain particular cases [17], cannot be recovered from the partial measurement.

Therefore, to enable the application of standard Mueller matrix analysis techniques (including parameterization, decomposition, and polarimetric imaging) it is essential to define appropriate criteria for completing $\mathbf{M}$. To do so, the following natural complementary criteria are proposed: (1) the measured 3×3 matrix must match the upper-left 3×3 submatrix of the synthesized 4×4 submatrix; (2) the completed 4×4 matrix should satisfy the four covariance conditions (i.e., the associated covariance matrix must be positive semidefinite), and (3) among the infinitely many Mueller matrices fulfilling the above conditions, $\mathbf{M}$ should be taken as the one that exhibits the highest degree of polarimetric purity.

The third criterion seeks to minimize the polarimetric randomness in the synthesized matrix. The nine known parameters exceed the seven required to describe a pure (i.e., single-component) system, but they are always consistent with systems involving two or more components. Among these, the two-component option (requiring up to twelve parameters) is the simplest choice besides allowing for maximal achievable purity and includes, as a particular case, the single-component (pure) Mueller matrix

This approach leaves three degrees of freedom available for adjustment beyond the nine fixed by the measurement.

As previously discussed, a two-component system corresponds to a covariance matrix with rank $\mathbf{H} = 2$ ($\lambda_2 = \lambda_3 = 0$, $\lambda_0 \geq \lambda_1 > 0$), which is equivalent to $P_2 = P_3 = 1$, so that the last two constituents of the characteristic decomposition (the most random ones) are effectively suppressed.

### 4. Synthesizing the last file and column of the Mueller matrix

It is important to emphasize that, given only a 3×3 partial Mueller matrix, the original full 4×4 Mueller matrix cannot, in general, be uniquely recovered. In fact, given a partial 3×3 measured matrix, there exist infinitely many physical completions (4×4 Mueller matrices) compatible with the same 3×3 block. Our goal is therefore not to retrieve the unknown 'true' Mueller matrix (an impossible task, except for pure Mueller matrices), but to synthesize one physically admissible completion that (i) matches the measured 3×3 submatrix, (ii) satisfies the covariance conditions, and (iii) corresponds to the least random statistical model compatible with the data, i.e., a two-component system with maximal polarimetric purity.

Up to the experimental error tolerance, we know that the measured 3×3 matrix is the upper-left 3×3 submatrix of a physically valid Mueller matrix $\mathbf{M}$. Consequently, the (unknown) covariance matrix $\mathbf{H}$ associated with $\mathbf{M}$ must be positive semidefinite. This implies that its conjugate matrix $\mathbf{H}^*$ is also positive semidefinite. Since any linear combination with positive coefficients of Hermitian positive semidefinite matrices is itself positive semidefinite, it follows that the real part of $\mathbf{H}$, given by $\operatorname{Re}\mathbf{H} = (\mathbf{H}^* + \mathbf{H})/2$, must also be positive semidefinite.

From Eq. (4), we see that $\operatorname{Re}\mathbf{H}$ is fully determined by the nine (known) elements of the measured 3×3 matrix together with the (unknown) element $m_{33}$. Therefore, there exists at least one (and generally infinitely many) value of $m_{33}$ for which $\operatorname{Re}\mathbf{H}$ is positive semidefinite.

This observation leads to the first step of the 4×4 matrix completion procedure, which consists of scanning numerically sequential values $x$ of $m_{33}$ and checking whether the resulting $\mathrm{Re}\mathbf{H}$ has nonnegative eigenvalues. For each value of $x$, the corresponding matrix is synthesized using

$$\mathrm{Re}\mathbf{H}(x) = \frac{1}{4}\begin{pmatrix} m_{00}+m_{01}+m_{10}+m_{11} & m_{02}+m_{12} & m_{20}+m_{21} & m_{22}+x \\ m_{02}+m_{12} & m_{00}-m_{01}+m_{10}-m_{11} & m_{22}-x & m_{20}-m_{21} \\ m_{20}+m_{21} & m_{22}-x & m_{00}+m_{01}-m_{10}-m_{11} & m_{02}-m_{12} \\ m_{22}+x & m_{20}-m_{21} & m_{02}-m_{12} & m_{00}-m_{01}-m_{10}+m_{11} \end{pmatrix}. \quad (10)$$

Then, among the set of positive and negative values of $x$ compatible with the nonnegativity of the eigenvalues of $\mathrm{Re}\mathbf{H}(x)$, the one that maximizes the associated degree of polarimetric purity $P_\Delta(x)$ (maximal purity criterion), is chosen, leading to the synthesis of $\mathrm{Re}\mathbf{H}$. Note that, from the expression for $P_\Delta$ in Eq. (3), such maximization corresponds to the maximal absolute value of $m_{33}$ compatible with the covariance conditions for $\mathrm{Re}\mathbf{H}(x)$.

The next step for the completion of the covariance matrix $\mathbf{H}$, is to identify appropriate values for the remaining six unknown elements of $\mathbf{H}$, namely $m_{03}, m_{13}, m_{23}, m_{30}, m_{31}, m_{32}$. To do so, we enforce the two-component condition, $\mathrm{rank}\,\mathbf{H} = 2$, on $\mathbf{H}$, which can be formulated as follows.

Let us perform numerically the diagonalization $\mathrm{Re}\mathbf{H} = \mathbf{Q}\mathbf{L}\mathbf{Q}^\mathrm{T}$, where $\mathbf{L} = \mathrm{diag}(l_0, l_1, l_2, l_3)$, with $l_0 \geq l_1 \geq l_2 \geq l_3 \geq 0$, $\mathbf{Q}$ being the orthogonal matrix that diagonalizes $\mathrm{Re}\mathbf{H}$. Then, let us build the matrix

$$\mathbf{K} = \begin{pmatrix} l_0 & -i\sqrt{l_0 l_1} & 0 & 0 \\ i\sqrt{l_0 l_1} & l_1 & 0 & 0 \\ 0 & 0 & l_2 & -i\sqrt{l_2 l_3} \\ 0 & 0 & i\sqrt{l_2 l_3} & l_3 \end{pmatrix}, \quad (11)$$

whose diagonalization is given by

$$\mathbf{\Lambda} = \mathbf{U}^\dagger \mathbf{K}\mathbf{U},$$

$$\mathbf{\Lambda} = \mathrm{diag}(l_0+l_1, l_2+l_3, 0, 0)$$

$$\mathbf{U} = \mathbf{W}\mathbf{V},\ \mathbf{W} = \begin{pmatrix} 1 & 0 & 0 & 0 \\ 0 & 0 & 1 & 0 \\ 0 & 1 & 0 & 0 \\ 0 & 0 & 0 & 1 \end{pmatrix},\ \mathbf{V} = \begin{pmatrix} a & -ib & 0 & 0 \\ b & ia & 0 & 0 \\ 0 & 0 & c & -id \\ 0 & 0 & d & ic \end{pmatrix}, \quad (12)$$

$$a = \sqrt{\frac{l_0}{l_0+l_1}},\ b = \sqrt{\frac{l_1}{l_0+l_1}},\ c = \sqrt{\frac{l_2}{l_2+l_3}},\ d = \sqrt{\frac{l_3}{l_2+l_3}},$$

where the pair of nonzero eigenvalues as well as the diagonalization matrix $\mathbf{U}$ are determined by the (nonnegative) eigenvalues $(l_0, l_1, l_2, l_3)$ of $\mathbf{L}$.

Then, since real and imaginary parts of a Hermitian matrix transform independently under unitary similarity transformations, the complete Hermitian matrix $\mathbf{H}_{rec}$ that matches the established criteria can be synthesized as $\mathbf{H}_{rec} = \mathbf{Q}\mathbf{U}\mathbf{\Lambda}\mathbf{U}^\dagger\mathbf{Q}^\mathrm{T}$. Then, the corresponding complete Mueller matrix is calculated through Eq. (5).

The proposed procedure is applicable to any measured 3×3 incomplete Mueller matrix and satisfies the criteria of compatibility, consistency and maximal polarimetric purity as established in Section 3. When the original measured 3×3 matrix derives from a pure Mueller matrix, the procedure naturally yields the same result as the method proposed by Ossikovski and Arteaga [16].

## 5. Practical implementation of the completion procedure

For clarity and reproducibility, we summarize here the completion algorithm in a step-by-step form. Given a normalized 3×3 Mueller block $\mathbf{A}$, the synthesis of a physically admissible 4×4 completion proceeds as follows:

(1) *Trial embedding*. Embed $\mathbf{A}$ into a 4×4 trial Mueller matrix $\mathbf{M}(x)$

$$\mathbf{M}(x) = \begin{pmatrix} m_{00} & m_{01} & m_{02} & 0 \\ m_{10} & m_{11} & m_{12} & 0 \\ m_{20} & m_{21} & m_{22} & 0 \\ 0 & 0 & 0 & x \end{pmatrix}, \quad (13)$$

with the unknown element $m_{33} = x$.

(2) *Covariance matrix and admissible values of x.* The numerical procedure scan trial values of $x$, and compute the corresponding covariance matrix $\mathbf{H}(x)$ from Eq. (4) and, from Eq. (10), its real part $\mathrm{Re}\mathbf{H}(x)$. This allows for determining the set of admissible values of $x$ for which $\mathrm{Re}\mathbf{H}(x)$ is positive semidefinite, i.e., all its eigenvalues are non-negative.

(3) *Maximal-purity choice of x*. From Eq. (3), it follows that the maximal degree of polarimetric purity $P_\Delta(x)$ of $\mathrm{Re}\mathbf{H}(x)$ corresponds to the admissible $x$ with highest absolute value, hereafter taken as the definitive $m_{33}$.

(4) *Two-component covariance matrix*. Diagonalize $\mathrm{Re}\mathbf{H}(x = m_{33})$ through the corresponding orthogonal similarity transformation $\mathbf{Q}^\mathrm{T}\mathrm{Re}\mathbf{H}\mathbf{Q} = \mathrm{diag}(l_0, l_1, l_2, l_3)$, with the choice $l_0 \geq l_1 \geq l_2 \geq l_3 \geq 0$. Through Eq. (11) compute the $\mathbf{K}$ matrix from the two largest eigenvalues and the corresponding matrices $\mathbf{\Lambda}$ and $\mathbf{U}$ as indicated in Eq. (12). This allows to generate the definitive two-component and maximal-purity covariance matrix $\mathbf{H}_{rec} = \mathbf{Q}\mathbf{U}\mathbf{\Lambda}\mathbf{U}^\dagger\mathbf{Q}^\mathrm{T}$.

(5) *Completed Mueller matrix*. Finally, through Eq. (5) obtain the completed Mueller matrix $\mathbf{M}_{rec}$. By construction, $\mathbf{M}_{rec}$ contains exactly the measured 3×3 block $\mathbf{A}$, satisfies the covariance conditions, and corresponds to a two-component system with maximal polarimetric purity compatible with the data.

This algorithmic description completely specifies the implementation of the completion procedure and can be directly translated into numerical code.

## 6. Illustrative examples

To illustrate the reconstruction procedure in a controlled situation, we start from a known normalized

Mueller matrix $\hat{\mathbf{M}}$ (Eq. (13) below). This matrix plays the role of a 'hidden ground-truth' system that is not assumed to be known in an actual partial polarimetry experiment.

We then mimic an incomplete measurement by discarding the last row and column of $\hat{\mathbf{M}}$ and keeping only its upper-left 3×3 block, denoted by $\mathbf{A}$ (Eq. (16) below). In other words, in the subsequent steps we act as if only $\mathbf{A}$ had been experimentally measured. If this 3×3 matrix is framed with zeros in the last row and column, the resulting 4×4 matrix does not exhibit the maximal compatible purity while, in general, violates the covariance conditions and is therefore unphysical.

The proposed completion algorithm is then applied to $\mathbf{A}$: we determine the value of $m_{33}$ that makes the real part of the covariance matrix positive semidefinite and maximizes the degree of polarimetric purity, and we synthesize the remaining unknown entries so that the associated covariance matrix has rank at most two. This yields the reconstructed Mueller matrix $\hat{\mathbf{M}}_{rec}$, which is a two-component system fully compatible with the measured 3×3 block $\mathbf{A}$, but not necessarily close to the original 'hidden' matrix $\hat{\mathbf{M}}$ in its unmeasured entries.

To do so, let us consider the following normalized Mueller matrix

$$\hat{\mathbf{M}} = \begin{pmatrix} 1.000 & 0.185 & 0.097 & 0.168 \\ 0.216 & 0.742 & 0.311 & 0.109 \\ 0.077 & -0.282 & 0.280 & 0.009 \\ 0.063 & 0.274 & 0.053 & 0.360 \end{pmatrix}, \quad (14)$$

which has been generated from a convex sum (parallel composition) of four pure nonnormal Mueller matrices each exhibiting specific retardance and diattenuation. Consequently, $\hat{\mathbf{M}}$ is depolarizing, with the following eigenvalues of its associated covariance matrix

$$\lambda_0 = 0.677, \; \lambda_1 = 0.281, \; \lambda_2 = 0.038, \; \lambda_3 = 0.004, \quad (15)$$

and corresponding IPP

$$P_1 = 0.396, \; P_2 = 0.882, \; P_3 = 0.984. \quad (16)$$

Let us now take the upper-left 3×3 submatrix $\mathbf{A}$ of $\hat{\mathbf{M}}$, which is obtained by removing the last row and column.

$$\mathbf{A} = \begin{pmatrix} 1.000 & 0.185 & 0.974 \\ 0.216 & 0.742 & 0.311 \\ 0.077 & -0.282 & 0.280 \end{pmatrix}, \quad (17)$$

Matrix $\mathbf{A}$, when framed with zeros in the last file and column, does not satisfy the covariance conditions. The calculation of $m_{33}$ through the proposed procedure gives $m_{33,rec} = 0.525$ (where the subindex $rec$ denotes for reconstructed), so that the extension of $\mathbf{A}$ to a 4×4 matrix with zeros in the last row and column except for such a value for $m_{33}$, leads to the completion of the real part of the synthesized covariance matrix $\mathbf{H}_{rec}$; the eigenvalues of $\text{Re}\,\mathbf{H}_{rec}$ being nonnegative. Then, the synthesized complete Mueller matrix takes the form

$$\hat{\mathbf{M}}_{rec} = \begin{pmatrix} 1.000 & 0.185 & 0.097 & -0.168 \\ 0.216 & 0.742 & 0.311 & -0.218 \\ 0.077 & -0.282 & 0.280 & -0.752 \\ -0.071 & -0.219 & 0.721 & 0.525 \end{pmatrix}, \quad (18)$$

which corresponds to a two-component system whose associated covariance matrix $\mathbf{H}_{rec}$ has the following eigenvalues

$$\lambda_{0,rec} = 0.917, \; \lambda_{1,rec} = 0.083, \; \lambda_{2,rec} = \lambda_{3,rec} = 0.000, \quad (19)$$

and corresponding IPP

$$P_{1,rec} = 0.834, \; P_{2,rec} = P_{3,rec} = 1. \quad (20)$$

The fact that the fourth row and column of $\hat{\mathbf{M}}$ (Eq. (13)) and $\hat{\mathbf{M}}_{rec}$ (Eq. (17)) differ significantly is therefore not a deficiency of the method but an unavoidable consequence of the information loss incurred when only the 3×3 block is measured. The reconstruction selects, among infinitely many admissible completions, the one with maximal polarimetric purity, which in general will not coincide with the unknown original matrix.

All numerical values reported in Eqs. (17)–(19) have been obtained by applying steps 1–5 described in Section 5, with no additional assumptions.

This example also illustrates the purification effect of the procedure, leading to a synthesized Mueller matrix, whose last row and column are replaced by compatible ones with the highest possible contribution to the degree of polarimetric purity. In polarimetric imaging, applying this completion pixel-wise to 3×3 measurements leads to an effective reduction of depolarization noise, analogous to isotropic depolarization filtering [40-42]. In practice, the increase in the IPP translates into enhanced contrast in the reconstructed polarimetric parameters, as has been experimentally demonstrated in Ref. [42]. In our framework, this 'filtering' effect naturally emerges from the two-component maximal-purity completion strategy, without requiring any a priori information about the sample.

As an additional theoretical example that illustrates the limiting situation where the polarimetric randomness is maximal, let us consider a perfect depolarizer $(P_\Delta = 0)$, whose normalized covariance and Mueller matrices are $\hat{\mathbf{H}}_{\Delta 0} = (1/4)\,\text{diag}(1,1,1,1)$ (with four equal nonzero eigenvalues) and $\hat{\mathbf{M}}_{\Delta 0} = \text{diag}(1,0,0,0)$. The corresponding measured 3×3 matrix would be $\mathbf{A} \approx \text{diag}(1,0,0)$. The procedure for the calculation of the last row and column of the synthesized Mueller matrix leads to

$$\hat{\mathbf{H}}_{rec} = \frac{1}{4}\begin{pmatrix} 1 & -i & i & 1 \\ i & 1 & -1 & i \\ -i & -1 & 1 & -i \\ 1 & -i & i & 1 \end{pmatrix}, \; \hat{\mathbf{M}}_{rec} = \begin{pmatrix} 1 & 0 & 0 & -1 \\ 0 & 0 & 0 & 0 \\ 0 & 0 & 0 & 0 \\ -1 & 0 & 0 & 1 \end{pmatrix}, \quad (21)$$

where the eigenvalues of $\hat{\mathbf{H}}_{rec}$ are $1/2, 1/2, 0, 0$, thus indicating that $\hat{\mathbf{M}}_{rec}$ represents an equiprobable incoherent mixture of two components.

In this case, the lost information about the original perfect depolarizer $\hat{\mathbf{M}}_{\Delta 0}$ leads to a synthesized system whose Mueller matrix $\hat{\mathbf{M}}_{rec}$ represents a perfect circular polarizer that differs substantially from $\hat{\mathbf{M}}_{\Delta 0}$. This does not prevent the applicability of the procedure but constitutes a limiting case of maximal possible distance between an (unknown) original Mueller matrix and the one synthesized from a partial 3×3 measurement. Obviously, when the operator of the polarimeter has extra information about the sample, the assumed criterion of maximal purity can be avoided and replaced by other appropriate criteria.

## 7. Conclusion

This work addresses the problem of interpreting incomplete Mueller matrices arising from partial polarimetric measurements in which only the upper-left 3×3 submatrix is accessible. A physically consistent and operationally justified procedure has been introduced to synthesize a compatible and appropriate full 4×4 Mueller matrix from such partial data. The proposed completion method is guided by three complementary criteria: (i) compatibility with the measured data, (ii) consistency with the covariance conditions, and (iii) maximal polarimetric purity.

The approach exploits the structure of the Hermitian covariance matrix associated with a Mueller matrix and identifies, among the infinitely many admissible completions, the one that corresponds to the least random (most pure) two-component statistical model. This synthesis strategy ensures the physical admissibility of the reconstructed matrix and allows its inclusion in conventional polarimetric processing and decomposition techniques.

The method has been validated through illustrative examples, demonstrating its robustness and physical relevance. In extreme cases, such as perfect depolarizers, the reconstructed Mueller matrix may significantly differ from the original one due to the intrinsic information loss.

It is important to emphasize that the objective of this method is not to precisely reconstruct the original matrix (an impossible task with insufficient data) but rather to find a compatible and appropriate Mueller matrix. This is achieved through a procedure designed to minimize the loss of relevant sample information while eliminating depolarization content to a certain extent, which can be interpreted as a reduction of polarimetric noise. Consequently, the reconstruction preserves the significant polarimetric information of the sample while enabling the use of a wide range of processing techniques typically reserved for fully measured systems. The procedure remains well-posed and provides a meaningful estimation, unless additional prior knowledge about the system is available, in which case alternative completion strategies can be adopted.

**Conflicts of interest:** The authors have nothing to disclose.

**Data availability**: No data were generated in this study.

**Funding**: This work was supported by the Ministerio de Ciencia e Innovación and Fondos FEDER (PID2024-156240OB-C22 and PDC2022-133332-C21) and Generalitat de Catalunya (2021SGR00138).

arXiv: 2510.06305v3 [physics.optics] (2025)